# Blue Ceramics: Co-designing Morphing Ceramics for Seagrass Meadow Restoration

**Rachel Arredondo**
Carnegie Mellon University
Pittsburgh, USA
rarredon@andrew.cmu.edu

**Ofri Dar**
Bezalel Academy of Art and Design
Jerusalem, Israel
ofrida95@gmail.com

**Kylon Chiang**
Carngeie Mellon University
Pittsburgh, USA
kylonc@andrew.cmu.edu

**Arielle Blonder**
Hebrew University of Jerusalem
Jerusalem, Israel
arielle.blonder@mail.huji.ac.il

**Lining Yao**
Carnegie Mellon University
Pittsburgh, USA
liningy@andrew.cmu.edu

**ABSTRACT**

Seagrass meadows are twice as efficient as forests at capturing and storing carbon, but over the last two decades they have been disappearing due to human activities. We take a nature-centered design approach using contextual inquiry and iterative participatory designs methods to consolidate knowledge from the marine and material sciences to industrial design. The sketches and renders documented evolved into the design and fabrication guidelines. This pictorial documents a dialogue between designers and scientists to design an ecological intervention using digital fabrication to manufacture morphing ceramics for seagrass meadow restoration.

**CSS Concepts**

• Human-centered computing~ Human computer interaction (HCI)~ Interactive systems and tools~ User interface toolkits

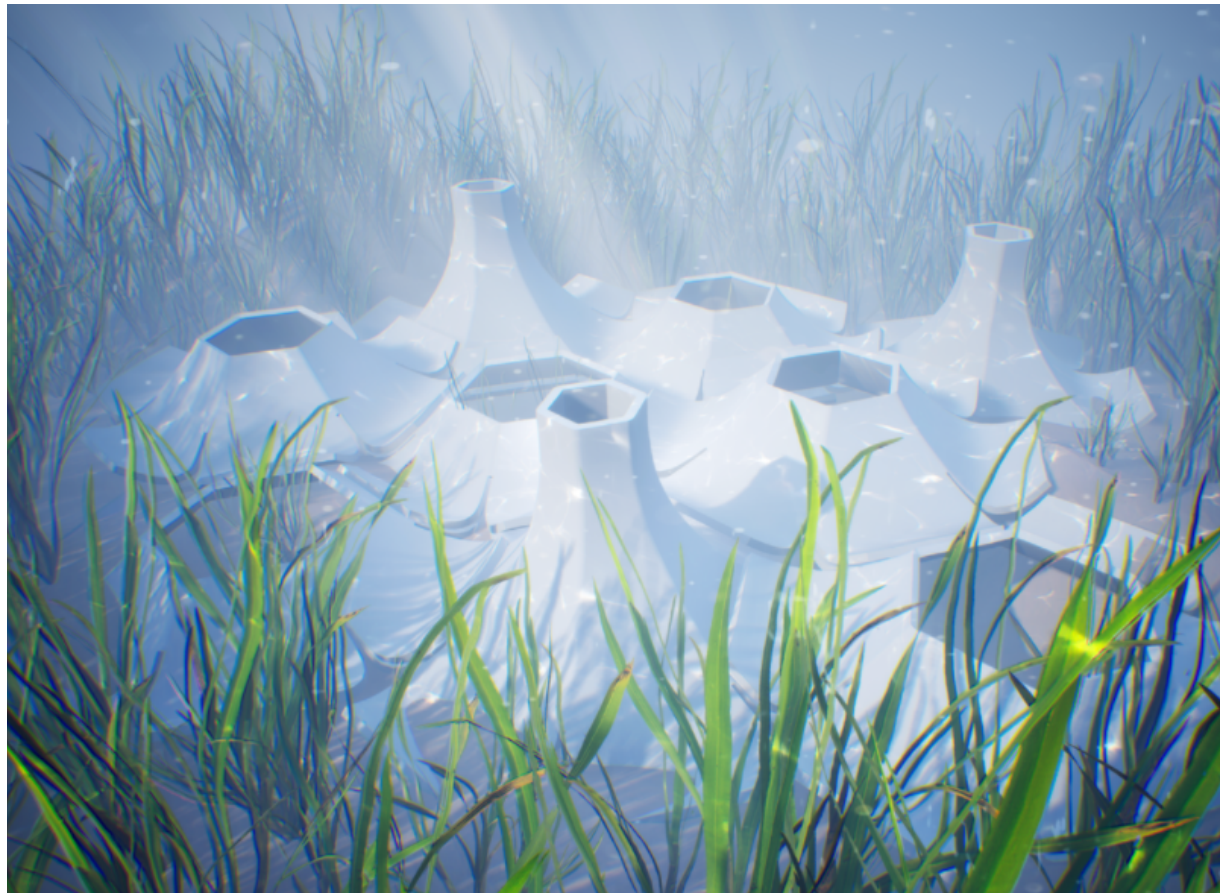

## INTRODUCTION

As the effects of climate change increase, seagrass meadows face challenges largely due to the impact of human activities in coastal marine environments [14]. Seagrass meadows are vital for healthy marine ecosystems [Fig. 1], as well a store large amounts of "Blue Carbon" [6]. Our nature-centered design approach [12] places focus on adapting human-centered design methods such contextual inquiry and participatory design for a natural stakeholder. With seagrass meadows decreasing [16], interdisciplinary work between design and science is needed to develop new restoration techniques. A key challenge in seagrass restoration is the lack of conservation tools for use in the restoration efforts [14]. In this pictorial, we show the mutual efforts between designers and marine scientists to develop new restoration tools to reduce challenges around seagrass conservation by approaching the environment as a stakeholder.

Our research and testing explored co-designing an ecological intervention, and fabricating this intervention with morphing ceramics. Through this process, we developed "Blue Ceramics" as a modular system of ceramic tiles for restoring seagrass meadow ecosystems. Our goal for this pictorial is to present how the design opportunities identified through participatory design inspire and inform the initial hardware development for ecological interventions. We hope to bring attention to how human-centered design methods can be used to explore opportunities that bring in humans as part of the ecosystem to create more sustainable and resilient futures for people and plants.

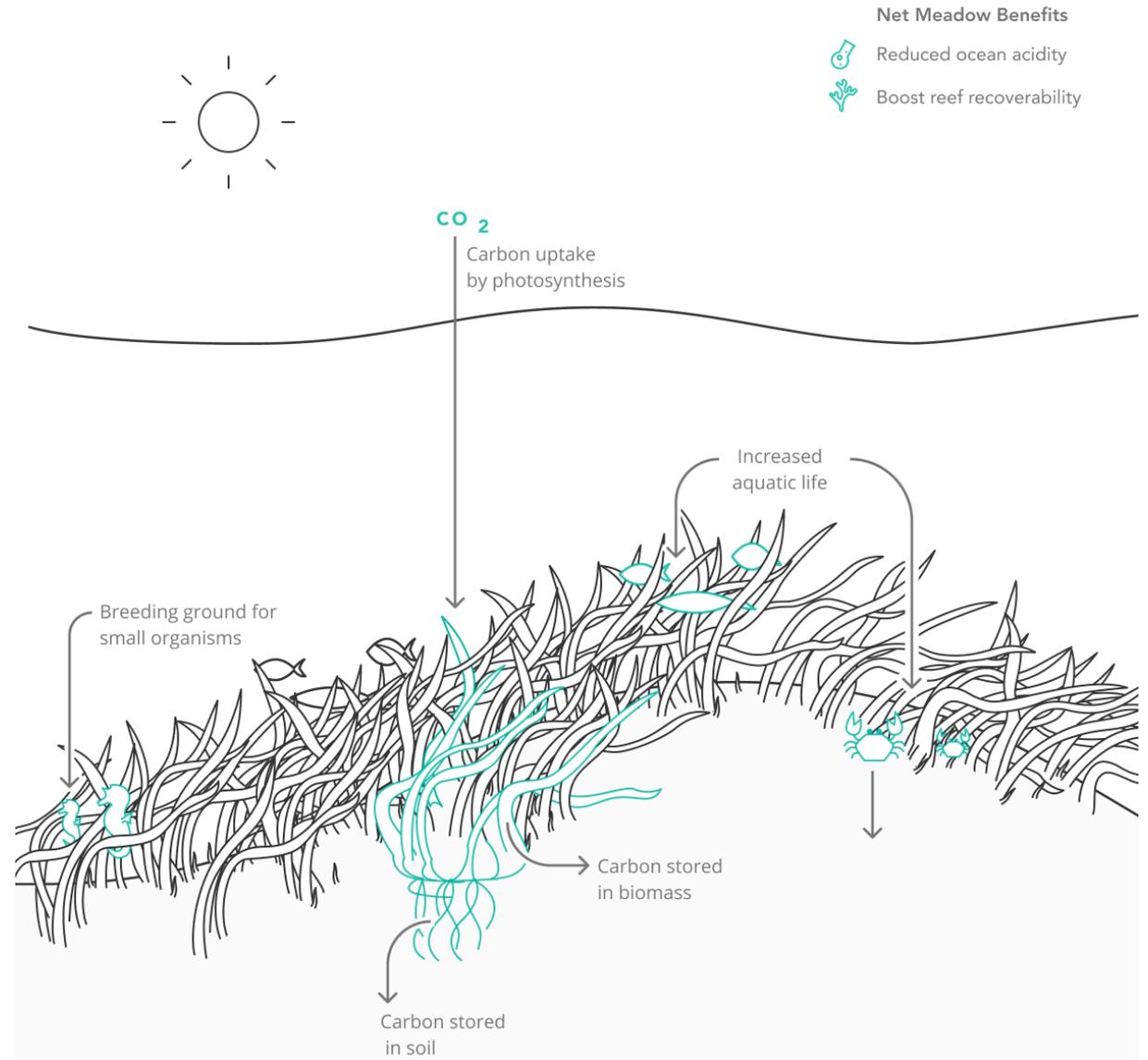

*Figure 1. Stable seagrass meadows are able to trap large amounts of Carbon in their roots and biomass. They also support commercially valuable species such as crabs, oysters, fish by providing breeding grounds and safety from predators.*

**MOTIVATION**

This project originally set out to explore the opportunities for ceramics in coral reef restoration, but early on we pivoted to seagrass. Coral reefs near seagrass meadows are more resilient to the effects of climate change [14], but little design attention focuses on the humble seagrass [5]. As the seagrass meadows decline, it's unknown how the animals living in them will adapt to the changes [16]. Restoring seagrass meadows is critical to the ocean ecosystem. However, current restoration methods are either too broad or too narrow for repairing areas of meadows that need replenishment [14]. We set out to solve this challenge by identifying potentially improved restoration solutions, and carried out the initial lab test with physical prototypes.

**DESIGNING FOR SEAGRASS**

Seagrasses play a key role in reducing the sediment in the water through their blade structure. The blades provide resistance against the current trapping sediment and increasing water clarity and the amount of sunlight they receive [3]. When a seagrass meadow system starts to decline, it usually happens in small patches. These are known as "cold-spots" and cause downstream affects such as increased ocean acidity, loss of breeding grounds for key species, and algal blooms [10]. Our intervention targets these cold-spots at different levels to reduce erosion and seed loss through sediment traps. This has the added benefit of providing structure for seagrass roots to regrow. The overall system is designed to increase seagrass meadow resiliency by stopping seed and plant loss at the leverage point, vital to restoration efforts. This provides a space to help facilitate seagrass meadow restoration by exploring ways to keep the sediment stable by providing an intermediary structure.

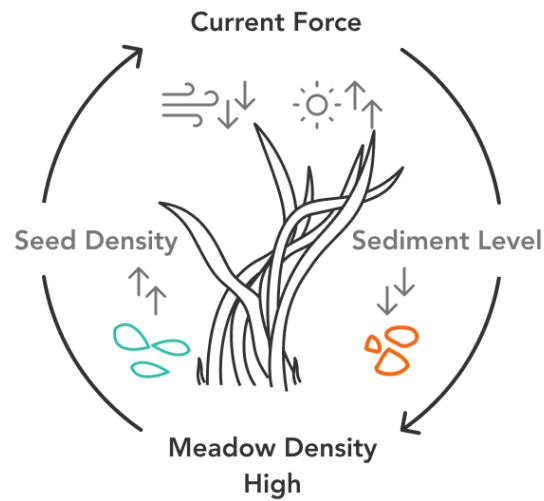

Figure 2. Higher meadow density increases current resistance reducing current flow

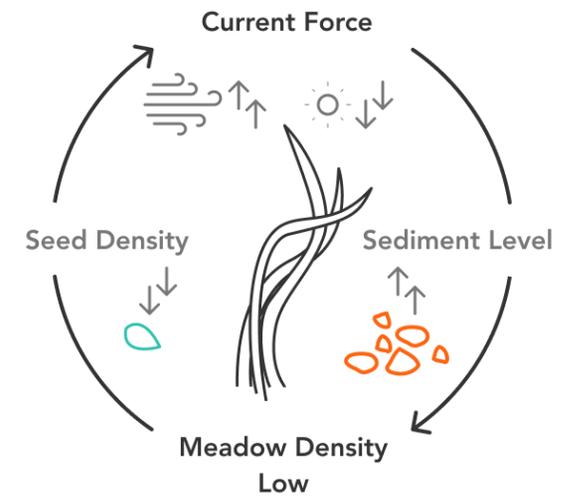

Figure 3. Lower meadow density provides less resistance increasing current flow

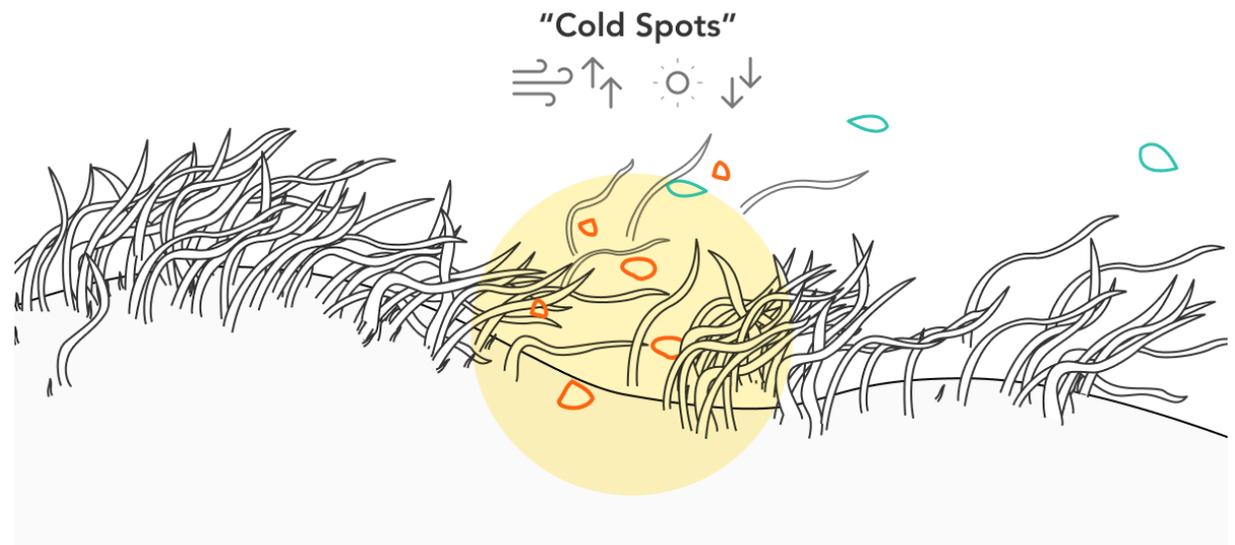

Figure 4. Cold spots reveal the sediment and seeds that was trapped by the seagrass roots. The increased currents remove these decreasing water clarity, and prohibiting regrowth.

**NATURE-CENTERED DESIGN**

Our initial contextual inquiry explorations to explored these expert's scientific and restoration knowledge through the lens of finding a design opportunity. We conducted semi-structured interviews with five experts in marine science and coastal restoration to understand the needs of seagrass meadow systems and restoration concerns. While we could not go into the field with them, we conducted semi-structured interviews with them via Zoom in their labs. We focused our interview protocol on prompting responses to pain points in their current work or research. The initial interviews with seagrass experts scoped our research to a few key points:

- Lack of restoration tools especially with the ability to target specific areas for restoration
- Water clarity is vital to ensuring good lighting conditions for seagrass meadow growth
- Increased sediment in the water from increased currents reduces water clarity
- Increased currents scatter seeds prohibiting natural regrowth

The initial semi-structured interviews also grounded our teams research into a particular type of seagrass, Eelgrass zostera marina, the most common seagrass in the Northern Hemisphere [13]. We also interviewed with experts in deploying similar systems for coral reefs also gave feedback on current pain points. These evolved into our guidelines for the intervention transport, assembly and deployment. We then adapted participatory designs methods to consolidate the knowledge from these interviews into design opportunity for an intervention centered on the real needs of seagrass meadows [Fig. 5].

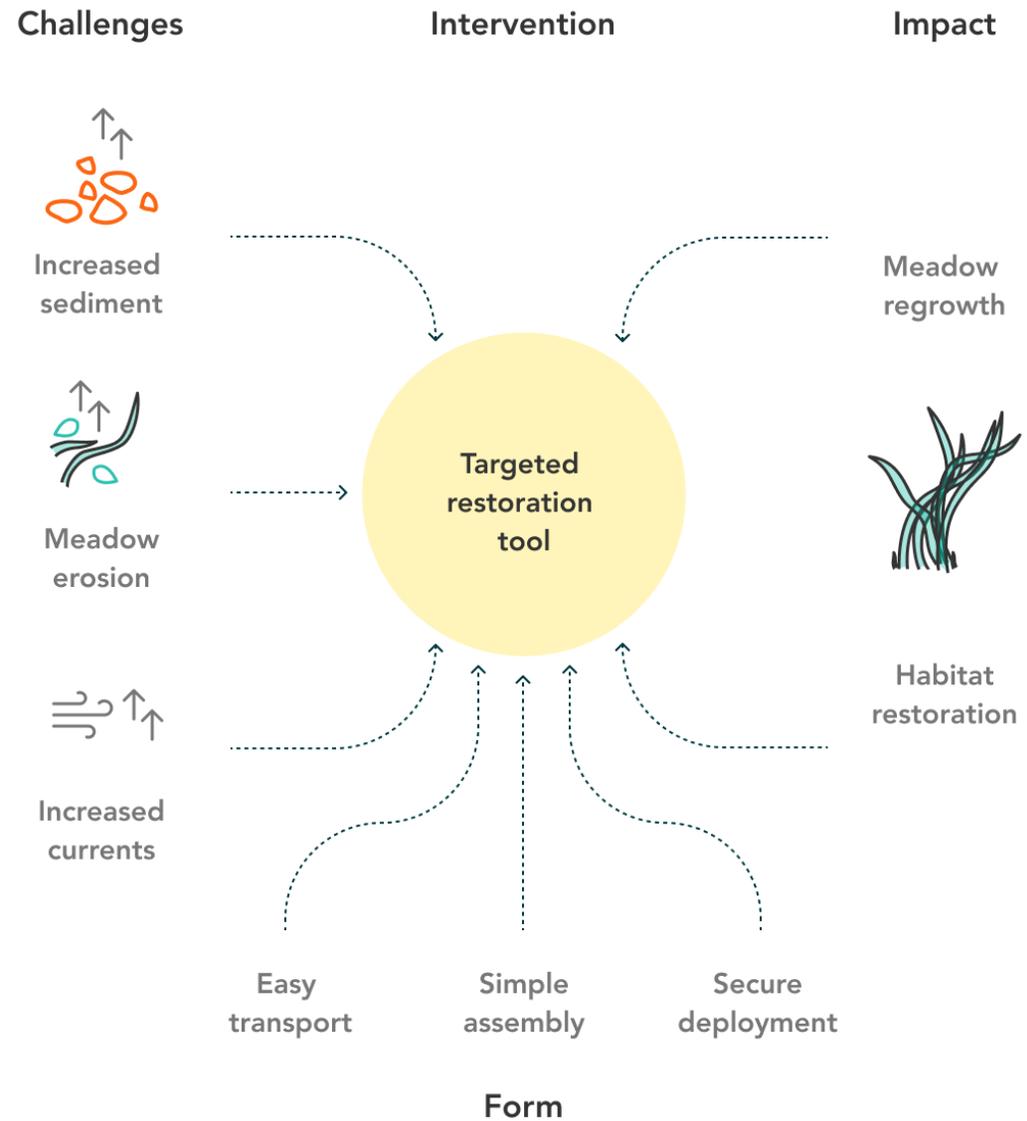

*Figure 5. Our research methods helped us scope our intervention design by providing guidelines for impact. This diagram shows how we integrated our knowledge into the final design.*

We paired with two different experts with experience in coastal restoration for Eelgrass (*zostera marina*) meadow systems to develop the intervention. From these sessions, we created our design challenge, intervention goals, and desired meadow impact. We adapted a collaborative sketching method from the "Homesense Kit Manual" by Alexandra Deschamps-Sonsino [4] to surface the key needs on the meadow. After a series of general prompts, the scientist identified which points to move forward with and the designer live sketched how these could be addressed. From this collaboration, we came up with a set of structures to be placed underwater that mimics how marshes and oyster beds help trap seeds and sediments [Fig. 7]. Another design idea focused promoting seed sedimentation and germination even under a strong current by providing a stable structure with holes to catch seed allow roots to anchor. These sketches served as the initial guidelines in exploring the intervention form.

## INTEGRATING THE RESEARCH INTO FORM

These methods helped us generate guidelines, rules, and principles about seagrass meadow systems and scope our work to a targeted design opportunity [Fig. 9]. We used the initial research and sketches as the basis for developing the form, and later the environmental simulation. These renders of the form design were used as visual artifacts in think-aloud tests with the marine scientists. We prototype several form designs and got feedback from experts before investing in the manufacturing of a particular shape [Fig. 8].

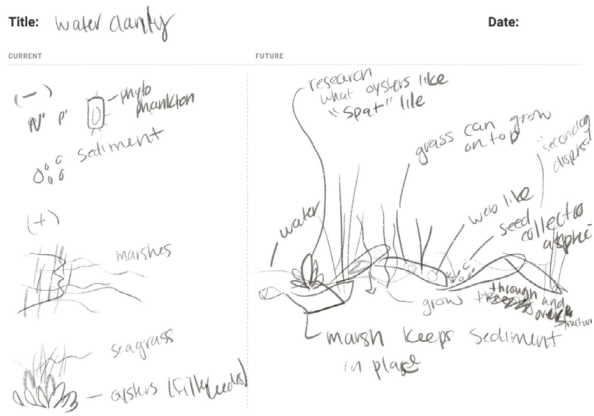

*Figure 6. This idea focuses on how marshes naturally trap sediment and seeds by sketching a tile network that mimics this ability.*

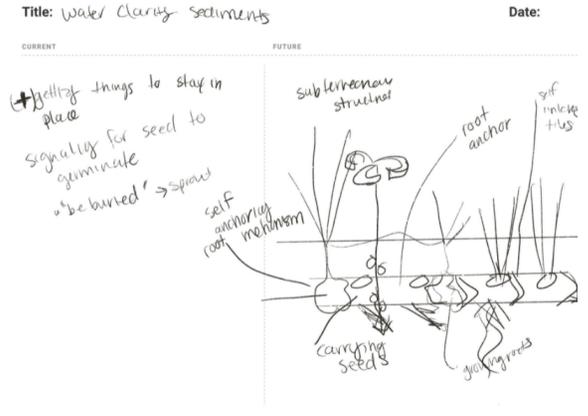

*Figure 7. This ideas focuses on how to provide a stable structure for seagrass roots in strong currents through an underground grid.*

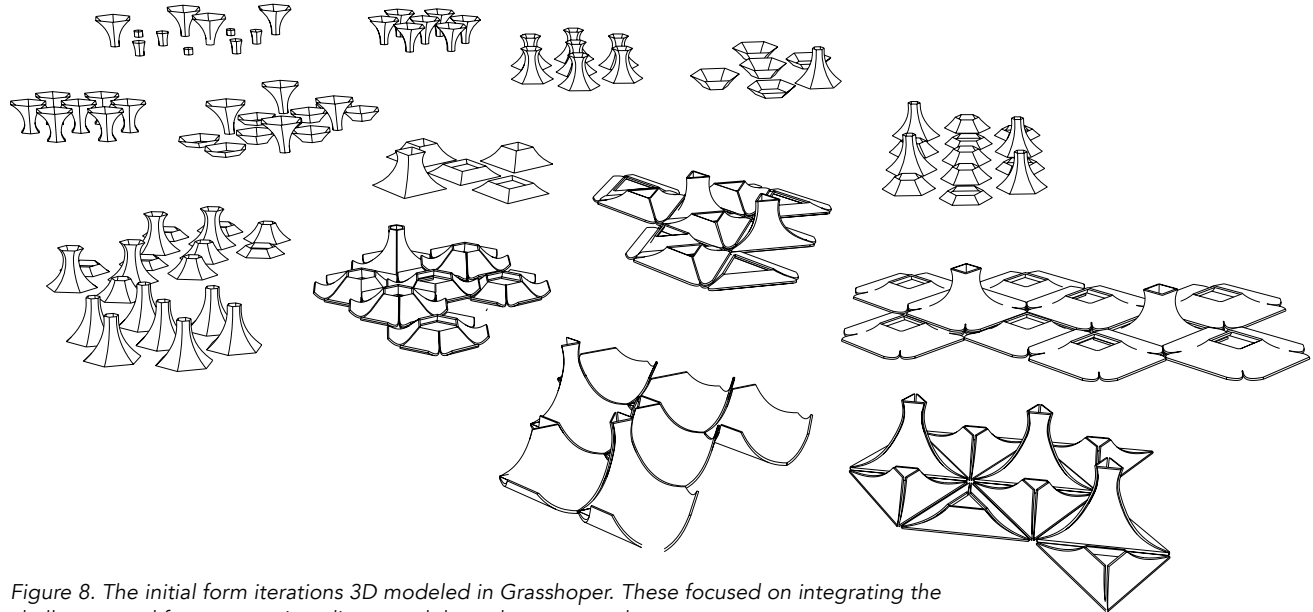

*Figure 8. The initial form iterations 3D modeled in Grasshoper. These focused on integrating the challenges and form constraints discovered through our research.*

*"I have been working with seagrass restoration for a long time and we didn't come up with anything like this. Just having something different to try is really exciting to me."*
— Coastal plant ecologist on our intervention design

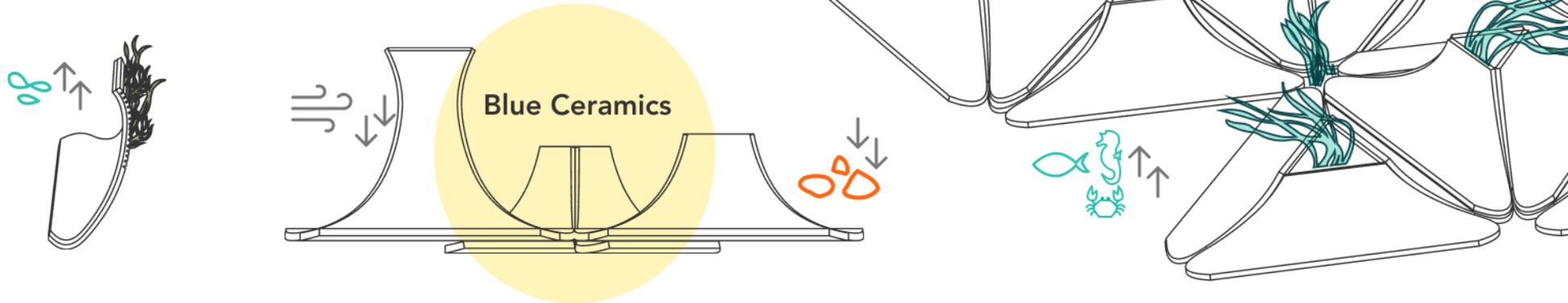

**Micro Level** — Add texture to increase seed anchoring

**Macro Level** — Increase topology to reduce current force. Coverage to prevent erosion and decrease sediment levels

**System Level** — Provide intermediate habitat for small organisms. Facillitate better conditions for meadow regrowth and restoration

*Figure 9. Blue Ceramics intervention design*

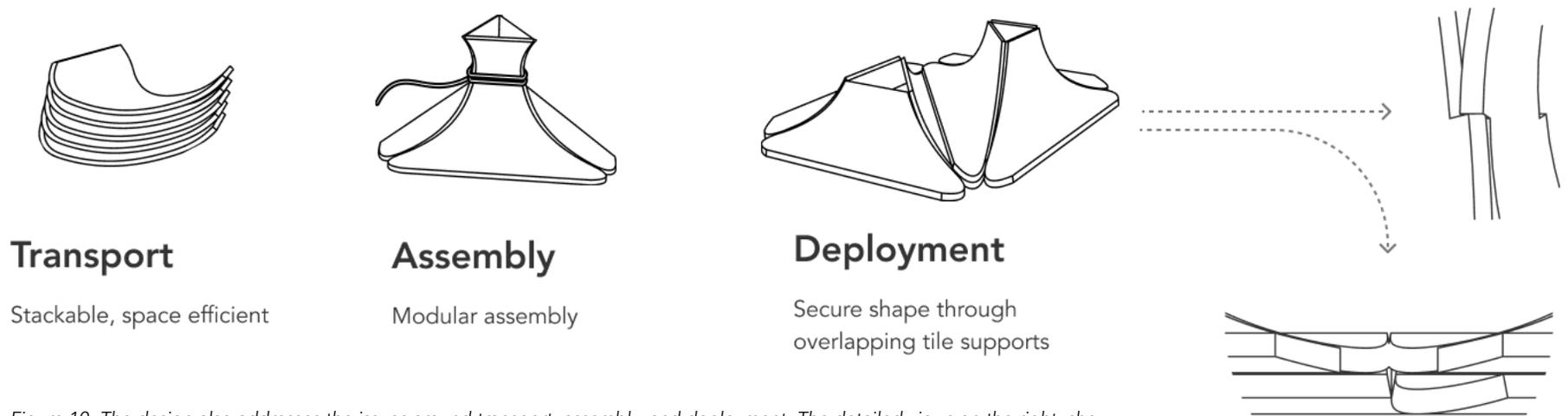

**Transport** — Stackable, space efficient

**Assembly** — Modular assembly

**Deployment** — Secure shape through overlapping tile supports

*Figure 10. The design also addresses the issues around transport, assembly, and deployment. The detailed views on the right, show how small overlaps placed in the structure help add stability to the form once deployed.*

Next we built an environmental simulation in Unreal. We used the systems information from our research and intervention design from the previous steps to build the 3D environment. This allowed us to virtually test out the deployment of the intervention—conveying the intervention goals, and form details to scientists, designers, and the general public alike. Additionally, the environmental simulation can be used to prototype deployment in different type of underwater environments.

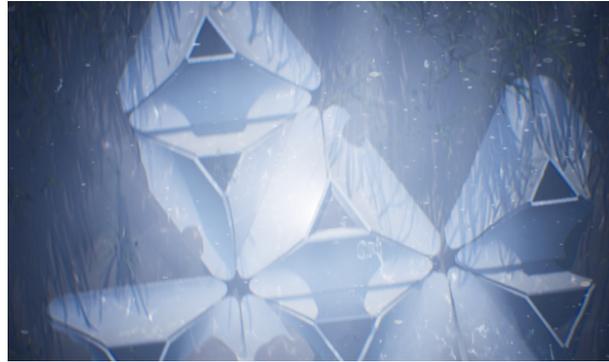 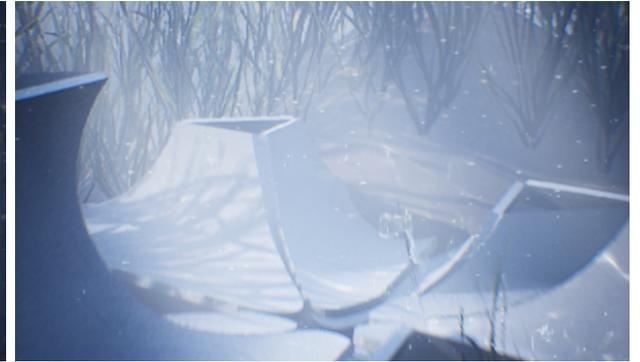

*Fig 11.  Blue Ceramics environmental render*     *Fig 12.  Blue Ceramics top view*     *Fig 13.  Blue Ceramics overlap view*

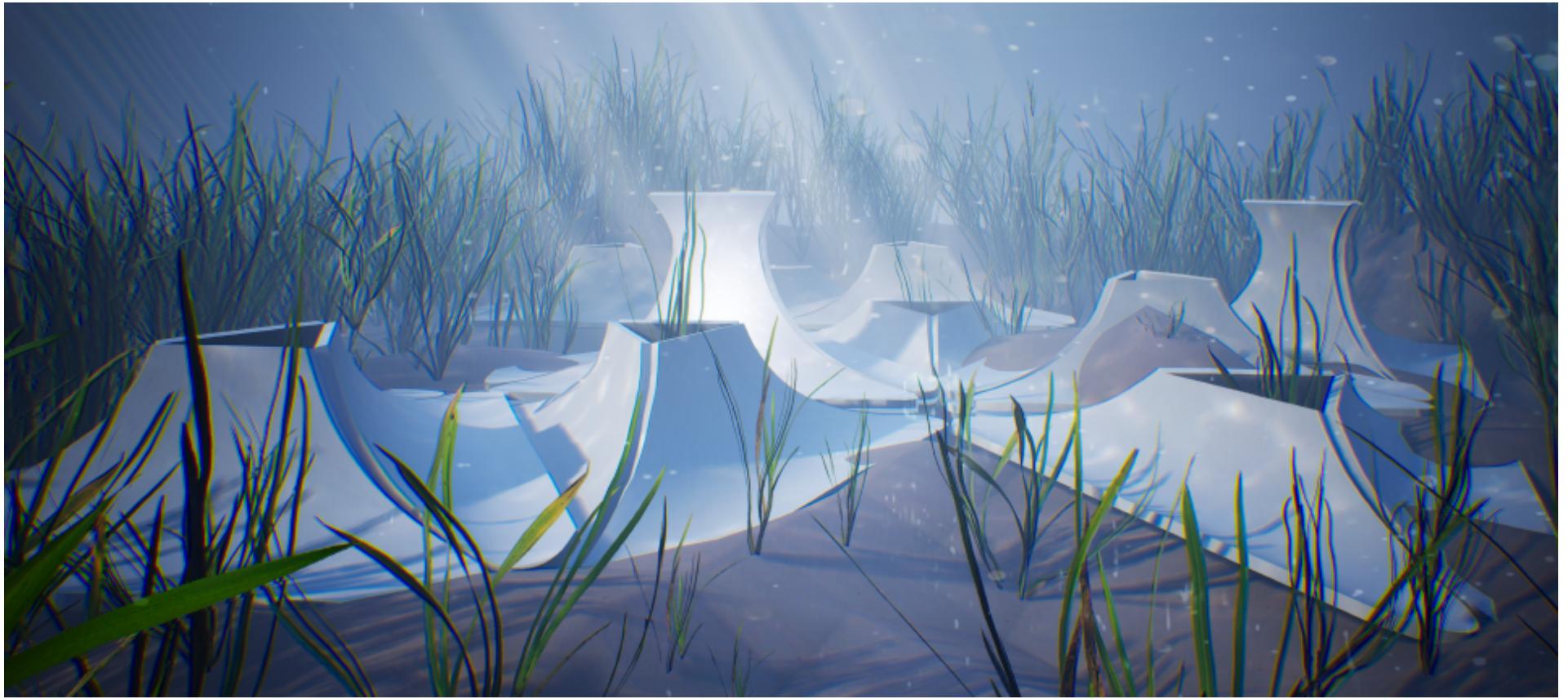

**DESIGN RATIONALE**

Our intervention design take into consideration the fabrication material and technologies involved in the manufacturing process. Ceramics is a natural material meaning we can leverage the fabrication and application, with its destruction within the same ecosystem [11]. For prototyping purposes, we use commercially available clay bodies to establish a material baseline of the morphing behavior of ceramics [Fig. 14-17], but the long term vision is to develop a "clay ink" from local sands and marsh clays. This ensures the intervention when the form breaks down through natural and environment stressors on the form it introduces no foreign materials.

Morphing materials with computational fabrication methods use less material and machine time and have a smaller carbon foot print than other methods [15]. We prototyped developing a morphing ceramics structures using two different fabrication techniques: CNC and 4D clay printing. Blue Ceramics is envisioned to be deployed in a variety of underwater locations with different environmental constraints. 4D printing provides a route for developing a fabrication process that can be modified easily to fit the specific needs of a specific seagrass meadow, such as height, width, and length [7].

**MANUFACTURING MORPHING CERAMICS**

Morphing ceramics take advantage of the tension created by two types of clay with differential shrinkage rates that are joined together then fired to create an isotropic, homogeneous positive reference Gaussian curvature [2, 8]. Our experiments used porcelain and stoneware, which our firing tests showed has a shrinkage rate of 13-15% and 7-8%, respectively. This contrast makes them the ideal candidates for making morphing bi-layer ceramic structures [8, Fig. 18].

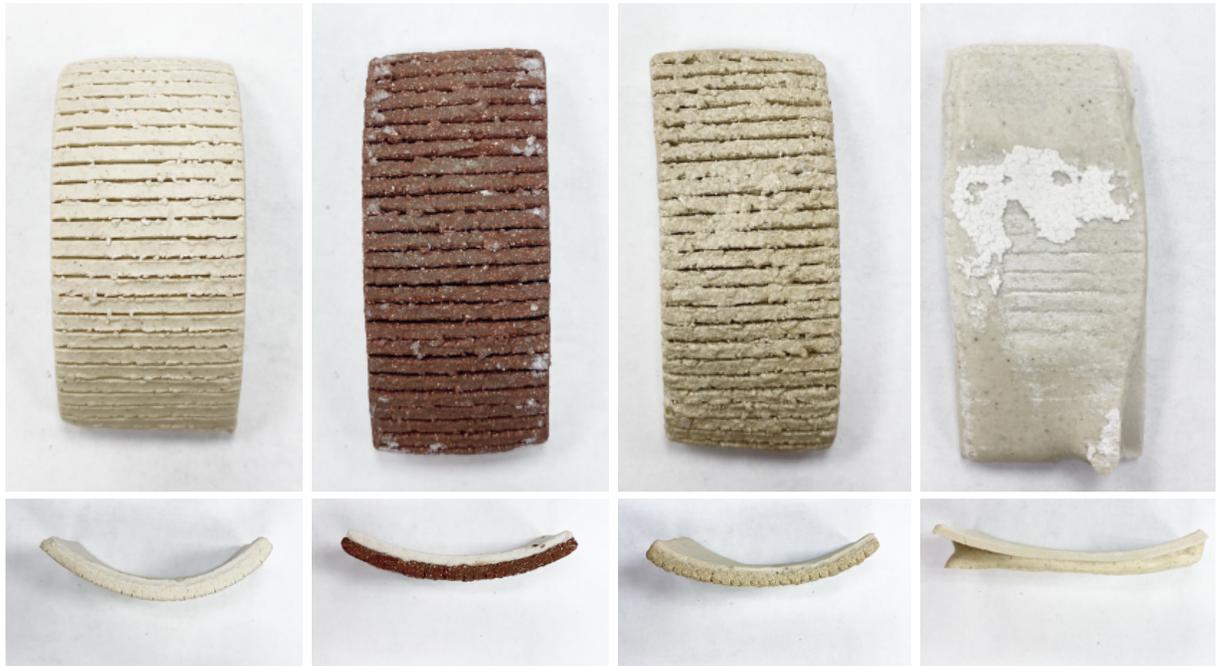

Figure 14. Stoneware with Talc exhibits the most highest stable morphing.

Figure 15. Red sculpture exhibits low but stable morphing.

Figure 16. Sculpture clay exhibits high but unstable but unstable morphing.

Figure 17. Stoneware without Talc vitrifies before morphing can be achieved.

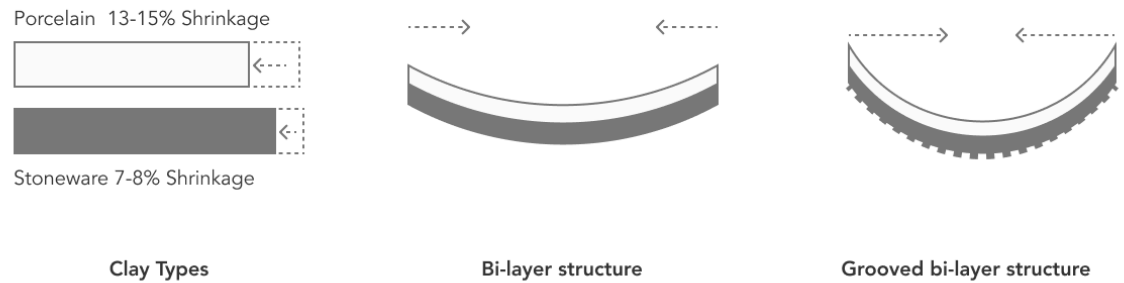

Figure 18. Our initial material tests established the shrinkage rates for our clays using a small test kiln. We used these tests to understand which clays would best be suited for morphing. We moved forward with testing the Stoneware with Talk.

We manufactured uniform test tiles using 2 fabrication methods, 4D printing[Fig. 19] and CNC grooving [Fig. 20], to explore the material properties and morphing behaviors of ceramics. These experiments focused on

1. Testing the mismatched stress in bi-layered ceramics
2. Observing the effects of geometric parameters such as grooving and material thickness on the resulting shape
3. Gathering data to derive and inverse design pipeline

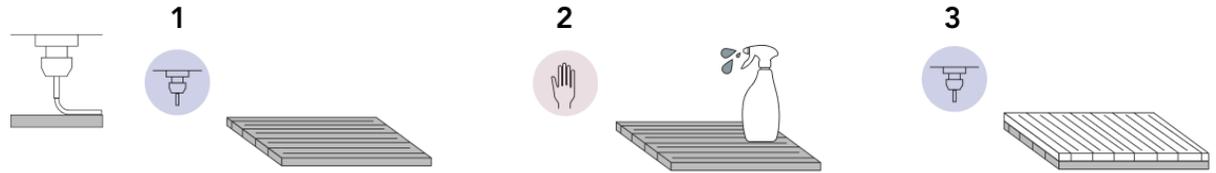

Figure 19. Using a Hyrel printer. 1. Load the printer with stoneware on a route aligned to the x-axis 2.Spray the with water 3. Load the printer with porclein aligned to y-axis.

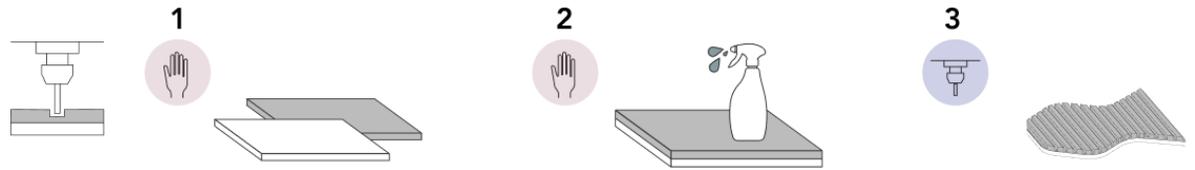

Figure 20. Using a CNC printer. 1. Hand fabricate the porcelain and stoneware slabs 2.Spray the with water 3. Groove the slabs

**Single Layer Tests**
In the initial print tests, we created single-layer tiles with identical dimensions but different print routes. While all samples shrank after firing, it was evident that using a print route perpendicular to the length-wise direction of the ceramic pieces causes the most shrinkage [Fig. 22]. This result later informed our manufacturing method for bi-layer samples.

**Bi-Layer Tests**
In the bi-layer tests, we combined clay sheets created through hand-forming, 3D printing, and CNC grooving to deepen our understanding of the relationship between manufacturing method and morphing behavior. The samples with the largest curvature had the following combinations:

- Printed porcelain with hand-formed stoneware
- Hand formed porcelain with printed stoneware
- Printed porcelain and stoneware printed perpendicularly to each other

This test provided further insight into the impact of print orientation on the resulting curvature [Fig. 21].

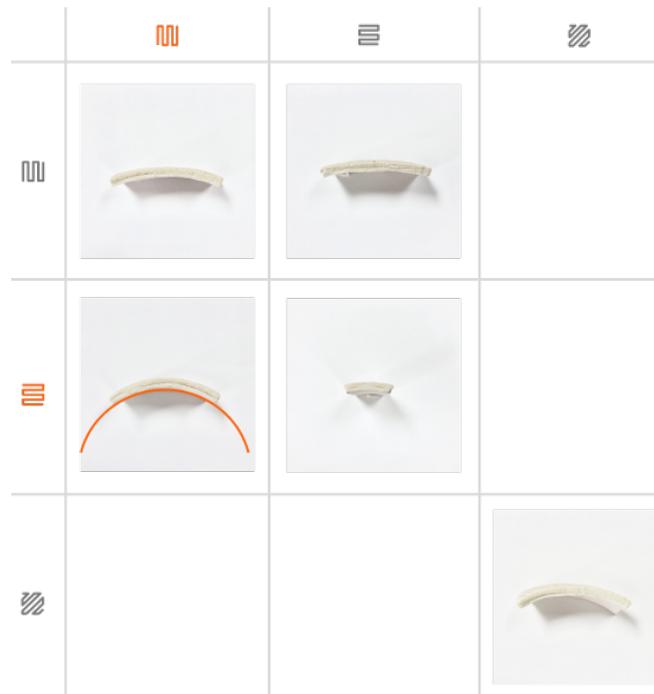

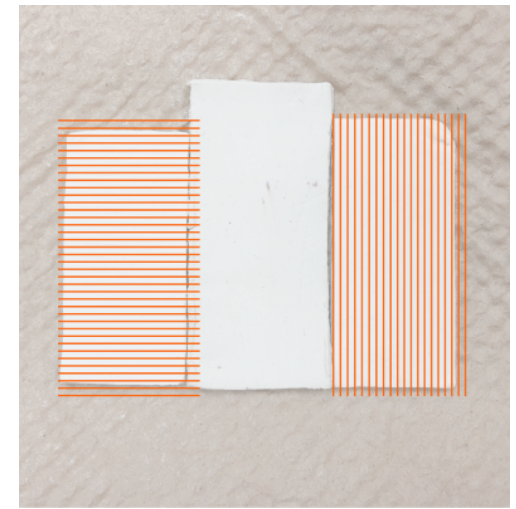

Figure 21. Left. Bi-layer layer testing of printing orientation

Figure 22. Above. Single layer testing of printing orientation

## Grooving Tests

We tested the effects of grooving the surface of clay sheets on the morphing process in parallel with single- and bi-layer tests. Earlier research demonstrated that surface grooves act as tension release points [9], which increases overall curvature. Our tests corroborated these results—samples with dense grooves greatly augmented the resulting morph [Fig. 23].

## Needle Adaptations

We also adapted the 3D printer as a grooving instrument by using an extruder head as a needle. By default, the printer holds the needle at 90°, which created rough lines and inaccurate curvatures. However, tilting the needle at a 45° angle creates clean grooves and more accurate curvatures by extension [Fig. 25.] A special adapter was later created to hold the needle in the correct position [Fig. 26].

Another design opportunity was revealed through grooving tests with round-tipped needles. This method creates ceramics with richly-textured surfaces, and when done with open-ended grooves, have the added ability to trap seeds and sediment [Fig. 27] We made small test samples verified this ability by making small test samples in the lab [Fig. 28]

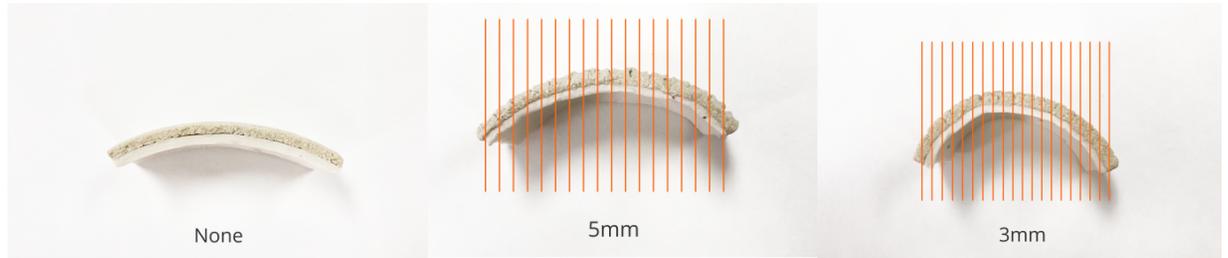

Figure 23. CNC machine grooving density tests

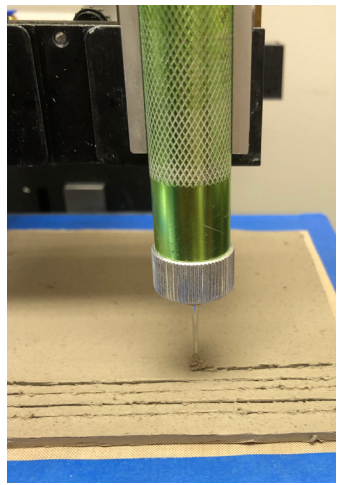

Figure 24. Hyrel printer with orginal printer head.

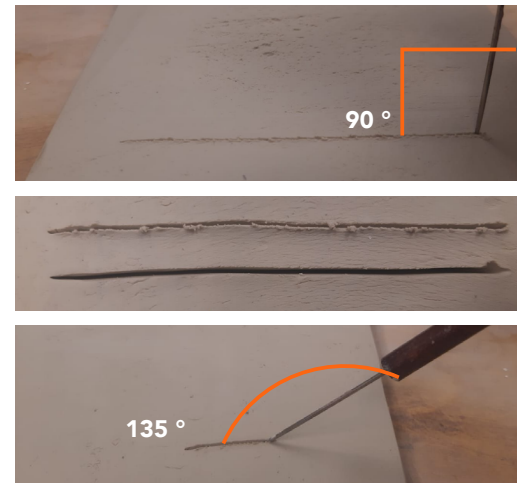

Figure 25. We noticed that while hand-grooving, we tilt the angle of the needle achieving cleaner edges.

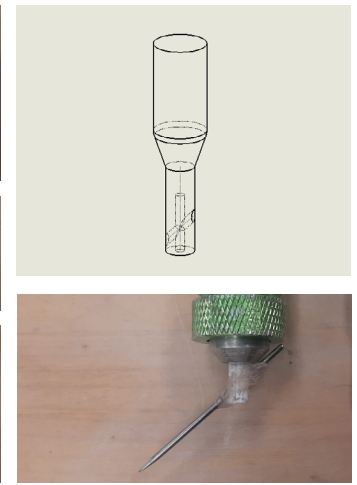

Figure 26. Adapted Hyrel printer head for angled grooving.

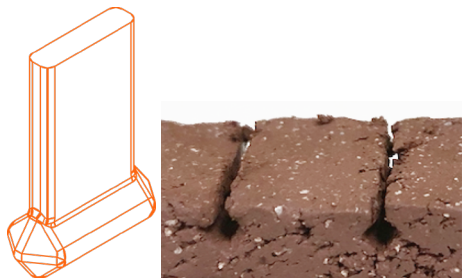

Figure 27. Round-tipped grooving design

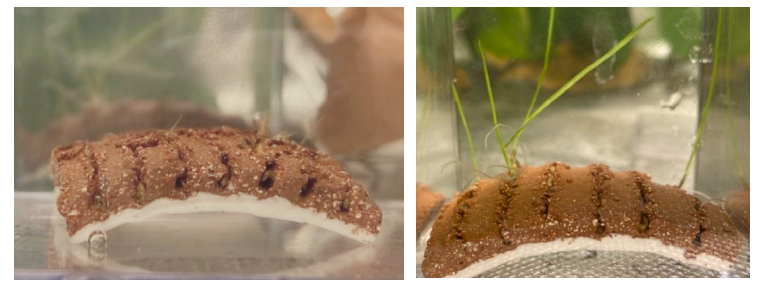

Figure 28. Round-tipped grooving design showing seeds growing on the tile.

### Inverse Design Pipeline

Our lab has developed a program that samples morphed pieces to systematically generate groove patterns that create the desired curvature on a given material [1]. We used this software during the later phases of our manufacturing process to increase the fidelity in relation to the design. We will continue to add tagged data to increase the accuracy of the program's output, and we hope that this will shift the burden of computation away from designers while leaving the freedom to imagine creative shapes [Fig. 29].

### Fabrication Tradeoffs

While additive manufacturing saves the efforts of creating an initially flat sheet clay, it can be time consuming. 3D printing clay is still slow to complete the printing of reasonably sized 3D objects [1]. However 4D printing uses less material overall and are geometrically easy to fabricate [14]. We envision a CNC grooving process on top of a flat sheet of clay through digitally controlled subtractive manufacturing. In combination with the inverse design pipeline, this process sets the stage for scaling the manufacturing process both in quantity and size with lowered cost of production and time.

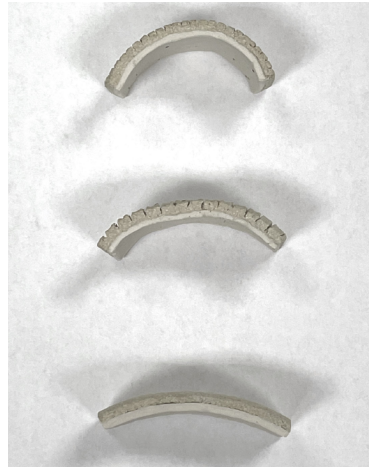
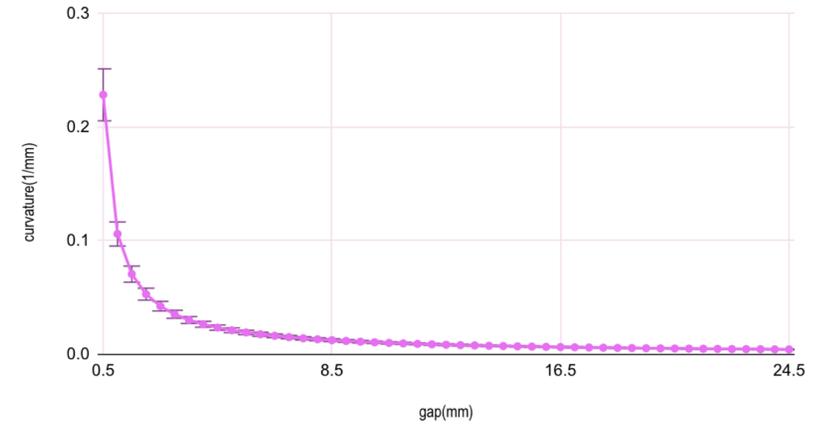

*Figure 29. Ceramic samples used to chart the relationship between the grooving and the morphing.*

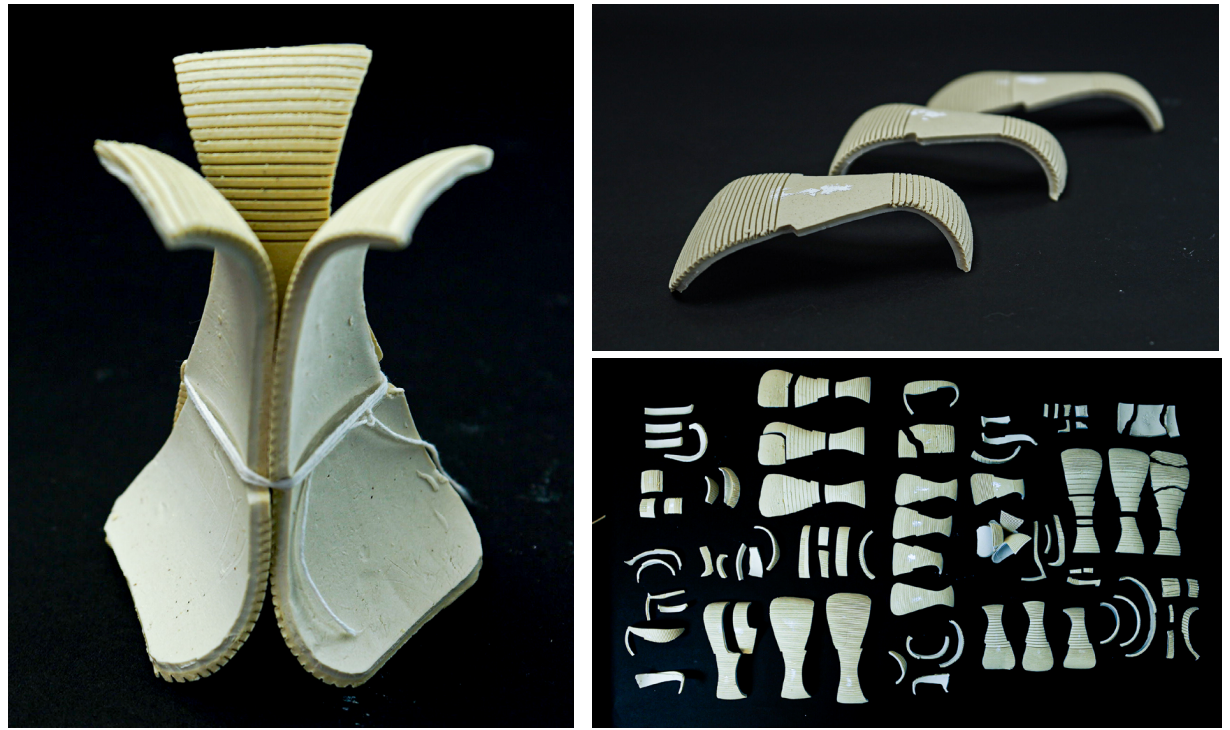

*Figure 30. Right. We constructed mid-sized morphing ceramic prototypes of the intervention design. Further testing is needed to build our understanding how the firing process affects morphing on larger tiles.*

## REFLECTIONS

Blue Ceramics is an ongoing project in the early stages, and this work will continue to evolve as more tests are completed. Future research should focus on thoroughly understanding the effects of clay material properties and firing temperatures on the morphing process. Once a repeatable baseline is established, then the design parameters can be refined. Early testing with small-scale prototypes validated our assumptions around the form and sediment interaction. [Fig. 31-32] Therefore, we chose to move forward with the current iteration of the form design, just at a larger scale for field testing.

From collaboratively sketching ideas to building morphing ceramic prototypes, our research brings a new lens on collaboration across disciplines by using familiar design methodologies to target nature instead of humans. This work demonstrates that the nature-centered design process proved to be the perfect framework to translate knowledge between designers and marine scientists to design ecological interventions. This pictorial highlights how our process brings together knowledge at the intersection of many different fields to design a new tool for seagrass meadow conservation. We hope this work broadens the conversation around the potential for design to bring about more sustainable futures.

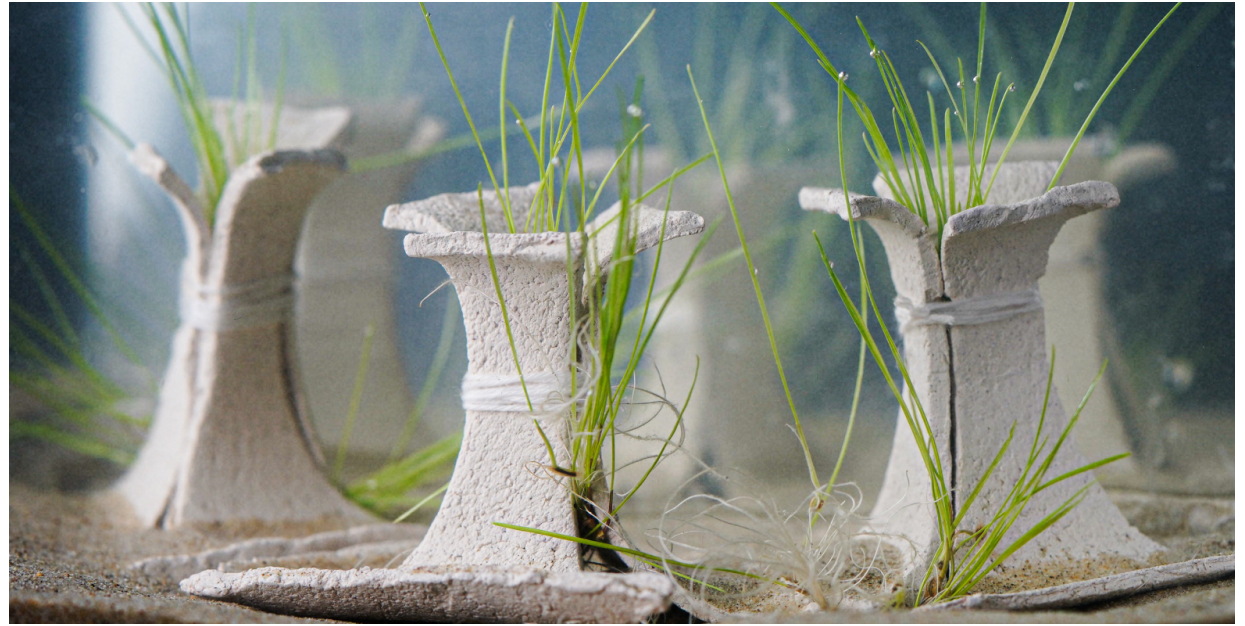

*Figure 31. Before building large scale prototypes, we designed smaller prototypes of the form design to test design details such as increasing tile overlap and stability.*

*Figure 32. Right. We placed these small form prototypes in a current tank and used different colored sands to understand current movement over the form. This was to understand how sediment of different sizes interacted on the form, not as a validation of the form itself which requires field testing.*

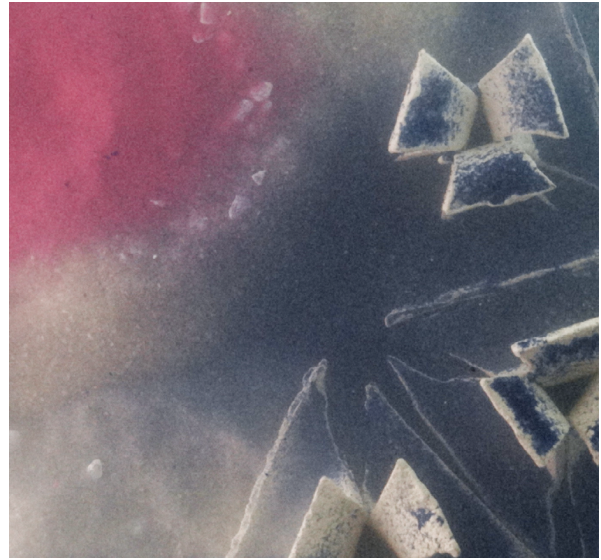
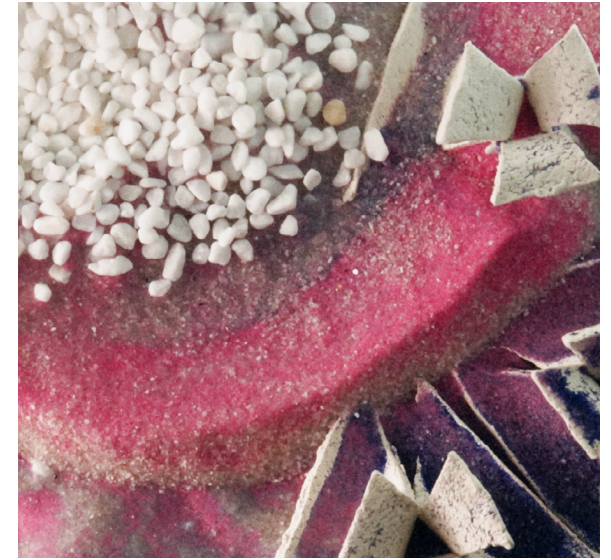

**ACKNOWLEDGMENTS**

This work could not have been completed with out the expertise of the following marine scientists: Dr. Jessie Jarvis, Jeffery Good, Dr. Abbey Engelman, Dr. Emmett Duffy.

The authors would like to thank Prof. Eran Sharon for his guidance on the physics of morphing materials. We thank Shira Shoval for her joint work in the creation of morphing ceramics.

We would would also like to thank Dan Vito, Donna Hendrick, and Bennet Graves from Fireborn Studios located in Pittsburgh, Pennsylvania for letting us use their space for our firing tests, and lending their expertise in kiln firing.

This project was funded by the Berkman Faculty Development Fund Grant at Carnegie Mellon University.